\documentclass[b5paper,twoside]{jpconf}
\usepackage{graphicx}
\usepackage{colortbl}
\usepackage{color}    
\usepackage{latexsym}                      

\begin{document}
 \def\red#1 {\textcolor{red}{#1}~} 

\title[HI Epoch of Reionization Arrays]{HI Epoch of Reionization Arrays}

\author[Greenhill \& Bernardi]{L. J. Greenhill \& G. Bernardi} 

\address{Center for Astrophysics, 60 Garden St, Cambridge, MA 02138 USA} 

\ead{greenhill@cfa.harvard.edu, gbernardi@cfa.harvard.edu}

\begin{abstract}
There are few data available with which to constrain the thermal history of the intergalactic medium  (IGM) following global recombination.  Thus far, most constraints flow from analyses of the Cosmic Microwave Background and optical spectroscopy along a few lines of sight.  However,  direct study of the IGM in emission or absorption against the CMB via the 1S hyperfine transition of Hydrogen would enable broad characterization thermal history and source populations.   New generations of radio arrays are in development to measure this line signature.  Bright foreground emission and  the complexity of instrument calibration models are significant hurdles.  How to optimize these is uncertain, resulting in a diversity in approaches.  We discuss recent limits on line brightness, array efforts including the new Large Aperture Experiment to Detect the Dark Ages (LEDA), and the next generation Hydrogen Reionization Array (HERA) concept.
\end{abstract}

\section{Introduction}

Characterization of the intergalactic medium (IGM) and early generations of  luminous compact objects before global reionization ($z>6$) is a frontier in observational cosmology.  For the first $\sim 100$\,Myr following  recombination and decoupling of radiation and matter, evolution  was dominated by gravity acting on diffuse distributions of dark and baryonic matter.  Models are robust owing to linearity of interaction.  For the remainder of the first $\sim 1$\,Gyr, nonlinear processes become increasingly relevant, e.g., with the widespread star formation, and understanding is limited.

There are no data with which to directly constrain evolution of the IGM prior to formation of luminous objects (a.k.a. the ``dark age,'' despite diffuse radiation).  Even during reionization of  small pockets initially and large expanses later, there are few data available to constrain the thermal history of the IGM,  source populations, and how these formed. An integrated global estimate of the redshift of reionization assuming a sudden transition ($z=11.0\pm1.4$) stems from measured large scale polarization of the cosmic microwave background (CMB, \cite{dun09}). Recent observations of the kinetic Sunyaev-Zeldovich effect on the CMB anisotropies have given a model dependent constrain on the duration of reionization $\Delta z < 7.9$ with reionization beginning at $z_{beg} < 13.1$ at 95\% CL (\cite{zah11}). 

Limits on the neutral Hydrogen (HI) column have been inferred from optical and near-infrared (NIR) spectra of luminous objects, where a trough blueward of  Ly$\alpha$  appears for $z>6.3$ (the ``Gunn-Peterson Trough'').  Neutral fraction is difficult to infer over a wide range of $z$ owing to the high absorption cross section and ready saturation.  However, inference may be drawn from modeling of the Ly$\alpha$ damping wing, including the effective radius of the proximate ionized zone.  As of this writing, only two quasars are known for $z > 6.4$ (\cite{fan06}), the most distant at $z\sim 7.09$ (\cite{mor11}).  Troughs have also been observed toward  gamma ray bursts, and there are 3 known for $z>6.4$, up to $z\sim 8.3$ (\cite{gre09}, \cite{sal09} and \cite{tan09}). 
Modeling of spectroscopic data for quasars and GRBs enables inference for a handful of lines of sight.  Broad sky coverage may be limited by nature -- there are relatively few  supermassive black holes that evolved in the first $\sim 1$\,Gyr -- and instrumental reasons -- wide-field NIR surveys and high sensitivity spectroscopic follow up is difficult (and subject to atmospheric windows.)

\begin{figure}[hb]
\begin{center}
\includegraphics{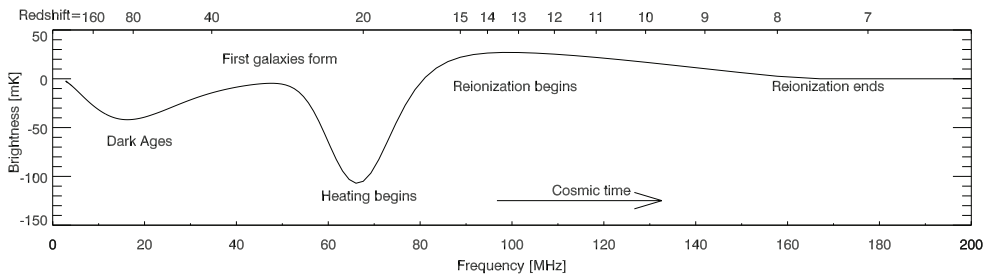}
\end{center}
\vspace{-0.2in}
\caption{\label{dc} \small Average sky brightness temperature  of the $\lambda$21\,cm line (i.e., spin temperature $\times$ optical depth)  up to the end of reionization \cite{pri10}.  Variation with redshift is driven by varying coupling between radiation (the CMB) and gas kinetic temperatures.}
\end{figure}

\begin{figure}[ht]
\vspace{-0.2in}
\begin{center}
\includegraphics[scale=1.0]{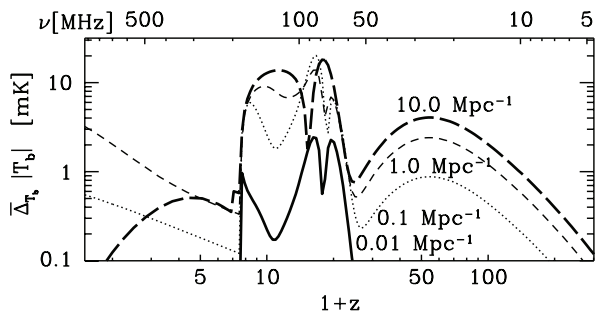}\hspace{1 pc}
\begin{minipage}[b]{8pc}
\caption{\label{ac}\small Power in fluctuations as a function of redshift for four spatial frequencies on the sky (adapted from \cite{pri11}). \\ \\ \\}
\end{minipage}
\end{center}
\vspace{-0.2in}
\end{figure}

The $\lambda$21\,cm transition of Hydrogen at high redshift has not yet been detected, but  as it would trace the diffuse IGM horizon to horizon without the need for illuminating compact objects, study of this transition has the potential to provide a new avenue by which to infer the evolution of the IGM and source populations.  This tracer complements CMB studies and optical/NIR spectroscopy and spectral-line origin in principle allows discrimination by redshift,  possibly into the Dark Age.   The physics of the $\lambda$21\,cm transition has been described in reviews (e.g., \cite{fur06}; \cite{pri11}, and references therein), but we summarize here salient points, referring to Figures\,\ref{dc} and \ref{ac}.  

Following recombination, the IGM cooled (adiabatically) faster than the CMB. 
When there was effective collisional coupling between $\lambda$21\,cm transition and gas temperature, the line will be seen in absorption against the CMB (both in the sense of a sky average and fluctuations).   This occurred for $z\gg 40$, when gas density was high, but before the first stars. Coupling faded with expansion of the Universe, reducing contrast between the HI line and CMB.  With early star formation, leakage of Lyman photons into the IGM recoupled the spin and kinetic temperatures via the Wouthysen-Field effect, resulting in a second absorption signature (e.g., \cite{pri10}).  The absorption depth and redshift of the minimum depend on the extent and evolution of heating sources  (e.g., star formation, X-rays from black holes, shocks), which may involve ``exotic'' physics (e.g., dark matter annihilation). 
Eventually, IGM heating drives the HI signature into emission before the HI component is largely destroyed; in this regime, the sky-averaged and  fluctuations signal will be more extended in fractional redshift and weaker in magnitude.

\section{Observations of the $\lambda$21\,cm line}
\label{instr_obs}

Sensitivity to the anticipated unpolarized mK signal will be limited by systematics.  These will arise principally from errors in instrument calibration and foreground models subtracted from the data. The two will be derived from similar data and likely to be coupled.  Unpolarized foregrounds will be smooth-spectrum mix of  extragalactic point sources and Galactic diffuse emission with angular scales overlapping those anticipated for the $\lambda$21\,cm signal (\cite{ber09}).  

Foreground brightness well exceeds the $\lambda$21\,cm signal.  The former rises  to {\it O}($10^3$)\,K at 100\, MHz even away from the Galactic plane (proportional to approximately $\nu^{-2.5}$ \cite{rog08}).  This is {\it O}($10^{4-5}$) times the $\lambda$21\,cm signal in sky-averaged power at  $\sim 70$ and 140\,MHz (Figure\,\ref{dc}). Fluctuations in foreground emission for ``quiet'' patches of sky are  {\it O}(1-10)\,K) (\cite{ber10}) and may be only  {\it O}($10^{2-4}$)  times the $\lambda$21\,cm fluctuations of  {\it O}(1-10)\,mK) at $\sim 140$\,MHz (Figure\,\ref{ac}).  The brightness of individual foreground sources extends down to the $\lambda$21\,cm signal.  This can be subtracted where it is above limits imposed by source confusion and contamination by sidelobes. However, this exceeds thermal noise limits for even modest integrations with arrays matched to the large angular scales of  fluctuations.  Below this nonthermal limit, other approaches are required (e.g., subtraction of an extant high-resolution sky or a statistical model).  The limit reported by \cite{ber10} is a few mJy at 150\,MHz and $\sim 2'$ resolution ($\sim$10\,K).
\begin{table}[t]
\vspace{-0.1in}
\begin{center}
\caption{\label{expt_tab} High Redshift $\lambda21$\,cm Experiments}
\centering
\small
\begin{tabular}{| l | c | c | c | c|c|r | l |}
\br
Expt.$^{\hspace{-0.03in}(1)}$ & Loc.$^{\hspace{-0.03in}(2)}$ & \multicolumn{2}{c |}{Elements} &  Area$^{(3)}$ & Bsln$^{(4)}$  & \multicolumn{2}{c|}{Band$^{(5)}$}\\
           & 	      &  Array         & Lone & (m$^2$)     & 	 (km)	    &      (MHz)      & {\it z}   \\
\mr

PAPER     & ZA &   128 & -- & 1000     & 0.2 & 110-180 & 6-12 \\
MWA       & AU &  128$\times$16 & -- & 3000     & 1  & 80-200 & 6-15 \\
\cellcolor[gray]{0.9}LEDA      & \cellcolor[gray]{0.9}US &  \cellcolor[gray]{0.9}256   & \cellcolor[gray]{0.9}4 & \cellcolor[gray]{0.9}3000/30  & \cellcolor[gray]{0.9}0.3 &  \cellcolor[gray]{0.9}30-88 & \cellcolor[gray]{0.9}17-42\\
\cellcolor[gray]{0.9}DAWN      & \cellcolor[gray]{0.9}US &  \cellcolor[gray]{0.9}256   & \cellcolor[gray]{0.9}--& \cellcolor[gray]{0.9}3000  & \cellcolor[gray]{0.9}0.1 &  \cellcolor[gray]{0.9}30-88 & \cellcolor[gray]{0.9}17-42\\
GMRT  & IN &  14 & -- & 8000     & 1    & 139-156 & 8-9\\
LOFAR & NL & 24$\times$768 & -- & 18000     & 3    & 120-200 & 6-10\\
\hline
HERA & c. 2016 & {\it O}($10^4$) & TBD & {\it O}(10$^5$) & {\it O}(3) & $<$100-200 & 6-12+  \\
\br
\end{tabular}
\end{center}
\vbox{\small $^{(1)}$ Shaded denotes  efforts to measure the angle-averaged HI spectrum.  LEDA and DAWN share the LWA1 aperture.  DAWN applies beam forming on cold patches of the sky.  LEDA instead cross-correlates the aperture ($\S$\ref{current_leda}).~~  
$^{(2)}$ IN: Pune, India; NL: Exloo, Netherlands; WA: Boolardy, Western Australia; ZA: Karoo, South Africa; US: VLA site, New Mexico.~~
$^{(3)}$ Approximate collecting area, mid-band, per polarization.~~ 
$^{(4)}$ Maximum array baseline. ~~
$^{(5)}$ Redshifts omit a 10\% guard band.  }
\vspace{-0.2in}
\end{table}

A more detailed study of foreground structure is still necessary for EoR observations.  At present, prediction of the unpolarized diffuse component requires interpolation between widely spaced frequencies.  Diffuse emission dominates the polarized sky (cf. pulsars and compact AGN).  There are no wide-angle low-frequency polarized sky surveys or consistent picture from the few patches that have been studied, and  angular and frequency structure in Stokes Q and U can differ considerably from that of Stokes I (\cite{ber09}).  

The point source population is somewhat better understood.  Catalogs of low-frequency point sources with arcminute resolution are available from the Cambridge surveys at 151-178\,MHz  (e.g., \cite{hal07}) and from the VLA 74\,MHz survey (\cite{coh07}). In the Southern Hemisphere, Culgoora Surveys at 80 and 160\,MHz (\cite{sle77}) provide measurements at arcmin resolution. A GMRT survey  that will provide finer resolution at 150\,MHz\footnote{http://tgss.ncra.tifr.res.in} is underway.

For single dipole instruments, a careful instrumental design coupled with a polynomial fit has been demonstrated to be effective in foreground subtraction (\cite{bow10}). However, uncertainty in antenna gain patterns (e.g., in response to the environment) and unmodeled responses of electronics have thus far limited sensitivity to the global 21cm signal other than for sudden reionization (\cite{bow10}).

For dipole arrays, a higher degree of calibration is possible, benefitting efforts to detect $\lambda$21\,cm fluctuations, or total power via array elements capable of measuring calibrated total power. Bright compact sources are used to calibrate direction and frequency dependent instrument response, and they may be subtracted directly from visibility data (\cite{tol07}, \cite{mit08}, \cite{ber10}, \cite{smi11}, \cite{kaz11}).  Foregrounds are mapped in parallel and, given a smooth and slow variation with frequency, galactic diffuse emission and sub-confusion level point sources can in principle be filtered (\cite{san05}, \cite{wan06}, \cite{jel08}, \cite{har09}, \cite{pet11}) provided that calibration errors do not result in pixel to pixel spectral fluctuations (e.g., \cite{bow08}).  Equivalently, calibration errors can couple a fraction of the polarized emission from Galactic sources into the unpolarized visibilities (as noted above), contaminating the $\lambda$21\,cm signal (\cite{ber10}, \cite{jel10}).  For idealized models of sky and instruments, it has been possible to filter out the polarization leakage by identifying the effects of different Faraday screens along various lines of sight and subsequently filtering the data (\cite{gei11}).  Active areas of investigation are in correction of ionospheric distortions (\cite{int09}), wide field polarization (\cite{ord10}), correction for antenna gain patterns (\cite{bha06},\cite{mor09}) and direction dependent deconvolution (\cite{pin10}, \cite{ber11}).  

\section{Ongoing $\lambda21$\,cm Experiments}
\label{current_exp}

\begin{figure}[t]
\begin{center}
\includegraphics[scale=0.80]{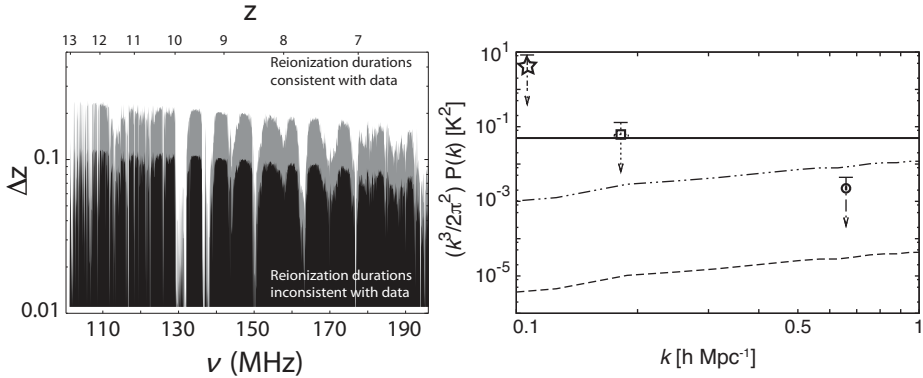}
\vspace{-0.15in}
\caption{\label{results} \small Experimental constraints on reionization thus far. {\it (left)}-- Lower limits for the breadth of reionization ($\Delta z$) for a ``sudden reionization'' model, estimated from single-dipole total-power measurements by EDGES (\cite{bow10}).  Shading denotes 68\% and 95\% C.L.  {\it (right)}-- Upper limits on 3D power spectra (z=8-9) from Westerbork data ($\star$) (\cite{ber10}), and the GMRT for two different foreground subtraction techniques (adapted from \cite{pac11}). Linear fits over 2.0\,($\Box$) and 0.5\,MHz ($\circ$) provide a foreground  subtraction model.  The 0.5 \,MHz limit may preclude a cold IGM at z$\sim$9 (dot-dashed line; \cite{jel08}).}
\end{center}
\vspace{-0.2in}
\end{figure}

Arrays thus far reflect a diversity of architectures, which is a consequence of incomplete knowledge of the low-frequency sky, complex coupling of dipole elements to the sky (and man-made transmissions), and uncertainty regarding optimization of foreground subtraction.  The RF hierarchy (i.e., layout of antennas) dictates the challenges to be met in signal processing and in calibration and imaging.  Equivalently, the available computing budget where signal processing is implemented with general purpose platforms (e.g., BlueGene, GPUs) constrains what RF hierarchies are practical.    We briefly review ongoing experiments (Table\,\ref{expt_tab}), emphasizing differences in architecture and signal processing.  Recent limits on the $\lambda$21\,cm signal are presented in Figure\,\ref{results}.

\subsection{Giant Metrewave Radio Telescope (GMRT)-EOR experiment}
\label{current_gmrt}

Fourteen GMRT dishes are located in a 1\,km$^2$ area, providing the low surface brightness sensitivity required to attempt detection of the $\lambda$21\,cm line.  A 16\,MHz wide 150\,MHz receiver band corresponds to z$\sim$8.5.  The narrow $\sim$3.3$^\circ$ field of view (FOV; half-power) of the dishes has enabled the experiment to begin without  wide-field polarization calibration and correction for ionospheric distortions.  The target field is centered on a pulsar and in a novel calibration scheme, gated visibilities (pulsar ON -- OFF) are differenced, thus eliminating the other (unswitched) foregrounds and  simplifying phase and polarization calibration (\cite{pen09}).   Foreground subtraction has thus far depended on fitting a frequency linear baseline on 0.5 and 2~MHz scales in visibility space, leveraging the anticipated smooth spectrum of the foregrounds (\cite{pac11}).

\subsection{Low Frequency Array (LOFAR)-EoR experiment}
\label{current_lofar}
Twenty-three high frequency (120-240\,MHz) LOFAR stations are arranged in the $\sim$3\,km core for use by the EOR experiment.  Each station comprises two clusters of 24 tiles of 16 dual polarization dipoles (768 total).  This is the deepest hierarchy of RF-element clustering for any EOR instrument.  Two levels of beam forming within the LOFAR stations enables good signal rejection but at the cost of complex sidelobe structure that may be difficult to map in assessment of coupling to the sky brightness distribution (cf. the GMRT).  Calibration and visibility plane foreground subtraction will be executed via generalized self-calibration techniques that incorporate direction dependent effects (\cite{noo04}, \cite{tol07}).  This is supported by stations far from the core, which are used to catalog point sources and to accumulate enough lines of sight to enable 3D tomography of the ionosphere (\cite{koo10}).  The EOR experiment is in its first year of data collection.

\subsection{Murchison Wide-field array (MWA)}
\label{current_mwa}

The MWA will comprise 128 tiles of 16 dual polarization dipoles, each with a beam former establishing a  single $\sim$20$^\circ$ half-power FOV at 150\,MHz.  EOR observing will utilize  a strongly centrally condensed pseudo-random configuration of $\sim 1$\,km$^2$ (\cite{lon09}).  Dense {\it u,v}-plane sampling and co-planarity will enable imaging via co-adding of warped snapshots (\cite{ord10}).  As such, wide-field correction can be achieved with geometric precision, and ionospheric distortion (approximated as a 2D rubber sheet)  can be implemented in image space.  The large number of baselines will enable detection of many calibrators per snapshot, a running estimation of the many, variable tile gain patterns, and correction during gridding of visibility data.   Source subtraction down to nearly the confusion limit will be accomplished via peeling in visibility space and image-based Forward Modeling will enable deconvolution  (\cite{pin10},\cite{ber11}).   Among EOR projects, so great reliance on image-based processing is unusual, but if successful, it may be relevant to much larger arrays, for which the volume of  of visibilities, $\propto${\it O}(N$_{\rm dipole}^2$), may preclude traditional techniques.

\subsection{Precison Array to Probe the Epoch of Reionization (PAPER)}
\label{current_paper}

PAPER comprises 64 broadband dipoles (100-200\,MHz) with plans to build out to 128 or 256 (\cite{par10}).  A shaped ground screen for each reduces what would otherwise be horizon-to-horizon dipole response to $\sim$60$^\circ$.  LOFAR, MWA, and PAPER represent a progression to flatter hierarchies vis-a-vis RF and signal processing architecture.   PAPER operates in a drift scan mode and benefits from the stability of the dipole gain patterns (cf. MWA and LOFAR).  The array can be reconfigured, enabling experimentation with pseudo random distributions for building a sky model  and redundant distributions that can benefit calibration and concentrate sensitivity on particular angular scales (\cite{par11}) as may be desired for first detection of $\lambda$21\,cm brightness fluctuations.

\subsection{Large-Aperture Experiment to Detect the Dark Ages (LEDA)}
\label{current_leda}

LEDA\footnote{http://www.ledatelescope.org} is distinct among arrays in that it will attempt detection of the $\lambda$21\,cm signal at $z \sim 20$ (30-88\,MHz band) at the end of the Dark Ages and thus to constrain initial conditions for reionization  (e.g., \cite{pri10}).   The target is  the angle-averaged signal, and a large-N array is used enable joint estimation of instrument calibration and a sky  model (in contrast to what is possible with a single dipole).  LEDA will attempt to leverage the {\it O}(10$\times$) magnitude of the $\lambda$21\,cm absorption trough at z$\sim$20 compared to emission at lower redshift.   A 512-input  full-Stokes, FPGA/GPU-enabled correlator  (e.g., \cite{cla11}) will be integrated into the 100$\times$110-m Long Wavelength Array (station 1) for 30-88\,MHz observing.  Outrigger dipoles will be instrumented for total-power measurement using calibrated noise sources.  Their gain patterns will be calibrated using cross-correlation data for baselines to the core stations.  The outriggers will be well separated from the core to reduce mutual coupling effects that contribute high-order terms in gain patterns, to provide baselines that resolve most diffuse Galactic emission, and to improve sensitivity to the point source population.

\section{Hydrogen Epoch of Reionization Array (HERA) }
\label{hera_pres}

The Hydrogen Epoch of Reionization Array (HERA) is a planned $2^{nd}$ generation (mid-decade) experiment that  targets  the detailed evolution of the $\lambda$21\,cm  power spectrum  (e.g., \cite{lidz07}), and to a lesser extent  imaging ionized ``bubbles''  around selected luminous early quasars \footnote{http://reionization.org/RFI2\_HERA.pdf.   This described a roadmap of instruments. Here we refer to the next, mid-decade experiment, dubbed HERA-2, simply as HERA.}.   In concept, HERA will have similar sensitivity to phase 1 of the Square Kilometer Array, though HERA will be an experiment optimized for study of the  $\lambda$21\,cm line  at $z>6$.  It will be timed to follow current generation US projects (PAPER, LEDA, MWA) and exploit lessons learned.

The HERA conceptual design comprises a centrally condensed, filled aperture with a  sparsely sampled ``halo'' of outlying antennas.  In seeking to characterize $\lambda$21\,cm power spectra at a detailed level, emphasis is on constraint of systematics during instrument design and in foreground characterization and removal.  The primary specifications affecting sensitivity are FOV, pass band, total collecting area, and area per element (Table\,\ref{expt_tab}).   The geographic distribution of elements enters vis-\`a-vis the quality of foreground mapping and source subtraction.   FOV and pass band dictate directly the range of $k$-space to which the array is sensitive.  For large $k$, i.e., fine angular scales, artifacts from subtraction of point-sources may become increasingly serious (\cite{dat10}).  Errors in reconstruction and subtraction of diffuse foreground emission, particularly in the presence of ionospheric fluctuations, will also reduce sensitivity over a range of $k$, although quantitative modeling of the effect has been limited.

\begin{figure}[ht]
\begin{center}
\vspace{-0.2in}
\includegraphics[scale=0.9]{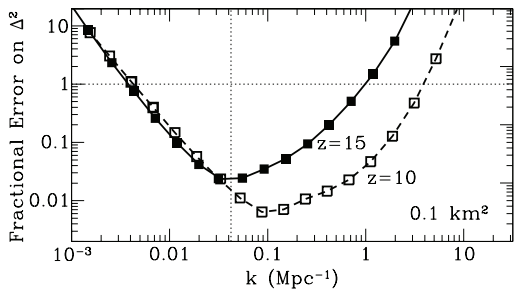}\hspace{0pc}
\begin{minipage}[b]{15pc}\caption{\label{sensitivity}\small \hspace{-0.05in}Indicative 3-D power spectrum measurement error vs a model, for a filled-aperture $10^5$\,m$^2$ array, assuming a neutral IGM, $1000^h$ integration, 8\.MHz bandwidth, ${\rm T}_{\rm sky}$ = (470, 1250)\,K, and 5000 antennas in a filled core and 3\.km diameter envelope. Symbols indicate independent bins. The vertical dotted line corresponds to bandwidth in $k$ units. The effects of calibration error and systematics are ignored.  Adapted from (\cite{fur09}).}
\end{minipage}
\end{center}
\vspace{-0.2in}
\end{figure}

The LOFAR architecture presents one model by which to accumulate collecting area (a sparse array with beam forming of many-dipole stations). In contrast, HERA is anticipated to use a  ``flatter'' array configuration.  This can be more computationally challenging, starting with cross-correlation, which is an {\it O}(N$^2$) problem.   For a collecting area of 10\,m$^2$ per antenna per polarization (more than for PAPER and LWA, and half that for MWA), $10^4$ elements are required,  for which correlation of 100\,MHz demands 40 PFlop\,s$^{-1}$.  A scalable solution for correlation drawing on GPU computing has been proposed by \cite{cla11}, with demonstration planned by LEDA.  Calibration carries a similar computational order (e.g., \cite{mit08}, \cite{edg10}), and scalable demonstration relevant to HERA may be achieved by LEDA and or MWA. 

{\bf Summary--} We have reviewed efforts to detect the high redshift $\lambda$21\,cm signal with radio arrays.  Diversity of approach  reflected in the design of first generation devices marks an ``epoch of experimentation''  from which strategies and optimizations for next generation arrays, such as HERA, will be based.

\end{document}